\documentclass[10pt,runningheads]{llncs}
\usepackage[ margin=2.67cm]{geometry}
\usepackage{graphicx}

\usepackage{enumitem}
\usepackage{svg}
\usepackage{epstopdf}
\usepackage[hidelinks]{hyperref}
\usepackage{booktabs}

\usepackage{amsmath,amsfonts}
\usepackage{algorithmic}
\usepackage{algorithm}
\usepackage{array}
\usepackage[caption=false,font=normalsize,labelfont=sf,textfont=sf]{subfig}
\usepackage{textcomp}
\usepackage{stfloats}
\usepackage{url}
\usepackage{verbatim}
\usepackage{graphicx}
\usepackage{cite}
\usepackage{amssymb}
\usepackage[final]{pdfpages} 
\usepackage{multirow}
\usepackage{threeparttable} 
\usepackage{marvosym}
\usepackage{tabularx}
\usepackage{pifont}

\usepackage{soul}
\usepackage{xspace}

\newcommand{\etal}{\textit{et al.}\xspace}
\newcommand{\blue}[1]{{\color{black}#1}}

\begin{document}

\title{TransURL: Improving Malicious URL Detection with Multi-layer Transformer Encoding and Multi-scale Pyramid Features}

\author{Ruitong Liu\inst{1,2} \and
Yanbin Wang\textsuperscript{\Letter}\inst{1,3} \and Zhenhao Guo\inst{3} \and
 Haitao Xu\textsuperscript{\Letter}\inst{3} \and
Zhan Qin\inst{3} \and
Wenrui Ma\inst{4} \and
Fan Zhang\inst{3}\thanks{Corresponding authors: Yanbin Wang and Haitao Xu}}

\authorrunning{R. Liu et al.}
%
\institute{Department of Engineering, Shenzhen MSU-BIT University, Shenzhen 518172, China\and Beijing University of Posts and Telecommunications, Beijing, 100876,  China\\ \email{liuruitong@bupt.edu.cn} \and
School of Cyber Science and Technology, College of Computer Science and Technology, Zhejiang University, Hangzhou, 310027, China
\and
College of Computer Science and Technology, Zhejiang Gongshang University, Hangzhou, 310027, China
\\
\email{wybpaper@gmail.com}}

\maketitle
\begin{abstract}
While machine learning progress is advancing the detection of malicious URLs, advanced Transformers applied to URLs face difficulties in extracting local information, character-level information, and structural relationships. To address these challenges, we propose a novel approach for malicious URL detection, named TransURL, that is implemented by co-training the character-aware Transformer with three feature modules—Multi-Layer Encoding, Multi-Scale Feature Learning, and Spatial Pyramid Attention. This special Transformer allows TransURL to extract embeddings that contain character-level information from URL token sequences, with three feature modules contributing to the fusion of multi-layer Transformer encodings and the capture of multi-scale local details and structural relationships. The proposed method is evaluated across several challenging scenarios, including class imbalance learning, multi-classification, cross-dataset testing, and adversarial sample attacks. The experimental results demonstrate a significant improvement compared to the best previous methods. For instance, it achieved a peak F1-score improvement of 40\% in class-imbalanced scenarios, and exceeded the best baseline result by 14.13\% in accuracy in adversarial attack scenarios. Additionally, we conduct a case study where our method accurately identifies all 30 active malicious web pages, whereas two pior SOTA methods miss 4 and 7 malicious web pages respectively. The codes and data are available at: \url{https://github.com/Vul-det/TransURL/}.

\end{abstract}



\section{Introduction}
Malicious URLs, systematically engineered by cybercriminals for illicit activities such as scams, phishing, spam, and malware distribution, constitute a significant cybersecurity risk. These URLs directly threaten user and organizational security, leading to privacy breaches, data theft, extortion, and compromising the integrity of devices and networks. Vade's 2023-Q3 report indicates a significant rise in phishing and malware, with malware volumes approaching a record high since Q4 2016, and phishing incidents increasing by 173\% from the previous quarter, reaching 493.2 million, the highest Q3 since 2015 \cite{Vade}.

In general, cybercriminals leverage deceptive hyperlinking as a key strategy in phishing schemes, often imitating credible entities such as Microsoft, Google, and Facebook \cite{cloudflare}. The ability to alter the display text of hyperlinks in HTML exacerbates the threat by camouflaging the true malicious nature of these URLs. This tactic poses a significant challenge to the detection of malicious URLs. 

\begin{figure}
    \centering
    \includegraphics[width=0.60\textwidth, height=0.12\textheight]{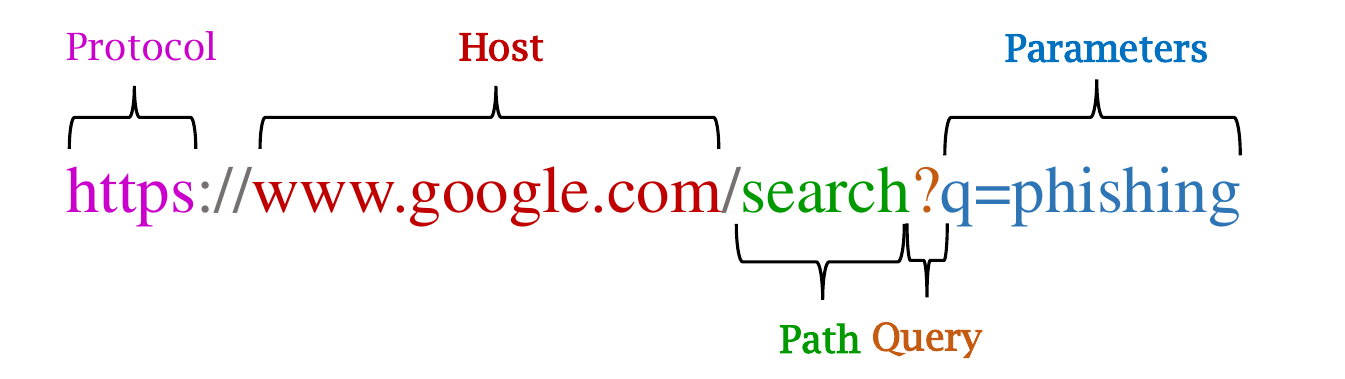}
    \caption{\textbf{Some major parts in a URL.}}
    \label{fig:fig1}
\end{figure}

Traditional detection methods like Phishtank and blacklist, heuristic, and rule-based approaches face delays and limitations in identifying new threats, as they depend on known URL structures and manual updates, struggling with novel malicious URLs \cite{sahoo2017malicious,li2020improving,mamun2016detecting,patgiri2023deepbf}. These limitations highlight the necessity for advanced machine learning techniques in the ever-evolving cybersecurity landscape.
On the other hand, earlier studies illustrates that malicious URLs display highly discernible string patterns(such as Fig. \ref{fig:fig1}) , including length, the quantity of dots, and specific words \cite{kim2022phishing,blum2010lexical}. These patterns play a vital role in threat analysis and provide the groundwork for training sophisticated classifiers. 

\begin{table*}
    \centering
    \caption{\textbf{Example of the BERT token sequence extraction from amazon web page.}}
    \begin{tabular}{|c|p{0.65\linewidth}|}
    \hline
        \textbf{URL} & \texttt{https://www.contactmailsupport.net/customer-service/amazon/} \\
    \hline
        \textbf{Token Sequence}  & \texttt{'[CLS]','https',':','/','/','www','.','contact','\string#\string#mail',' \string#\string#su','\string#\string#pp', '\string#\string#ort','.','net','/','customer','-','service', 
        '/','am', '\string#\string#az','\string#\string#on','/','[SEP]'} \\
    \hline
    \end{tabular}
    \label{tab:token}
\end{table*}
The advancement of deep learning has significantly propelled the development of malicious URL detection systems \cite{korkmaz2021phishing,maneriker2021urltran,chang2021research,moarref2023mc, de2023intrusion}, with Convolutional Neural Networks (CNNs) being the cornerstone in previous efforts, as exemplified by URLNet \cite{le2018urlnet}, TException \cite{tajaddodianfar2020texception}, and GramBeddings \cite{bozkir2023grambeddings}, which remain among the SOTA methods for malicious URL detection. However, the inherent technical constraints of CNNs have increasingly made it challenging to achieve substantial improvements in CNN-based malicious URL detection models. 

Recently, pretrained Transformer frameworks like BERT \blue{(Bidirectional Encoder Representations from Transformers)}\cite{devlin2018bert} have expanded their exceptional sequence modeling capabilities beyond natural language processing into various domains \cite{radford2019language, Brown2020language,da2024survey}. This innovative computational architecture and training paradigm offer enhanced contextual learning capabilities and function as purely data-driven end-to-end models (Table \ref{tab:token} provides an example of a BERT token sequence generated from a URL). However, the standard Transformer encounters specific challenges in malicious URL detection: 1) It struggles to capture character-level information due to its token-based input mechanism, critical for identifying subtle alterations in URLs. 2) Transformers are less effective than CNNs at detecting local patterns crucial for identifying potentially malicious substructures in URLs. 3)  Transformers lack the capability to directly discern the hierarchical structure inherent in URLs.

This paper introduces TransURL, addressing the challenges faced by Transformers in malicious URL detection tasks. TransURL, built on a specialized Transformer architecture, leverages embeddings from all encoding layers, integrating advanced multi-scale feature learning with spatial pyramid attention mechanisms to achieve state-of-the-art malicious URL detection.

The main contributions of this paper are as follows:
\begin{itemize}
\setlength{\itemsep}{0pt}
\item The proposed method achieves SOTA \blue{(State of the Art)} performance across a range of challenging scenarios, including class imbalance, small sample learning, multi-classification, cross-dataset validation, and adversarial sample attacks, comparing previous best methods. Furthermore, its practicality is further demonstrated through case studies.
\item The method introduced operates on a character-perceptive Transformer structure, effectively deriving embeddings that contain both subword and character-level information from a token sequence within a URL, all without relying on manual dual-input configuration.
\item Our method is the first to dynamically fuse multiple encoding layers of a deep Transformer framework, achieving nuanced multi-level feature extraction from URL sequences, and it provides empirical proof of the performance improvements attributed to this fusion of information.
\item We propose a joint training framework that combines the Transformer with multi-scale convolution and spatial pyramid attention techniques, leveraging their respective advantages. This innovative framework is capable of long-distance sequence modeling, supporting advanced multi-scale local feature extraction and global information aggregation.

\item Our rigorous experimental setup exposes that even the previously most effective methods have their limitations in certain scenarios, providing a vital testing framework for constructing models designed for practical use.

\end{itemize}

The paper unfolds as follows: Section \ref{sec:related} conducts a literature review, while Section \ref{sec:data} outlines the datasets used. In Section \ref{sec:method}, we provide a thorough explanation of the architecture and key components within our model. \ref{sec:experiments} details extensive experiments on malicious URL detection, benchmarking against baseline methods. A case study is then provided in Section \ref{sec:case}, and our conclusions are summarized in Section \ref{sec:conclusion}.

\section{Related Work}
\label{sec:related}
Malicious URL detection has a long-standing history in research. In this paper, we primarily review some recent works relevant to our study, which can be categorized into two types: CNN-based approaches and Transformer-based approaches.

\subsection{CNN-based Detection}
Huang et al. \cite{huang2019phishing} proposed a network that incorporates convolutional layers and two capsule network layers to learn the embedding representations of URLs. Wang et al. \cite{wang2019bidirectional} combined CNNs and RNNs to extract key features for measuring content similarity, integrating these with static lexical features extracted from URLs using Word2Vec for their detection model. URLNet \cite{le2018urlnet} introduced a dual-channel CNN approach for learning both character and word-level embeddings, combining these at the model's top. This method not only achieved state-of-the-art performance at the time but also inspired numerous subsequent studies \cite{tajaddodianfar2020texception,wang2022tcurl,hussain2023cnn,zheng2022hdp}, which all adopted the dual-channel feature extraction concept of URLNet. Recently, Bozkir et al. \cite{bozkir2023grambeddings} developed GramBeddings, a neural network that effectively combines CNNs, LSTMs \blue{(Long Short-Term Memory)}, and attention mechanisms. This network represents URL features through n-grams and has shown performance that surpasses URLNet in certain aspects.

The application of traditional neural networks in this field has seen widespread adoption. However, recent research suggests that their performance appears to have reached a plateau, leaving limited room for further improvement. Moreover, although these methods have advanced malicious URL detection, they still rely on manually initialized features at different levels (characters, words, or n-grams). In contrast, our proposed method is purely data-driven, requiring no manual engineering, and ingeniously implements feature extraction at both subword and character levels.

\begin{table*}[t]
    \centering
    \begin{threeparttable} 
        \captionsetup{position=bottom}
        \caption{\textbf{The statistical analysis of our datasets.}}
        \begin{tabular}{lllllllllllll}
            \toprule
            \textbf{Dataset}& \multicolumn{3}{l}{\textbf{Sample Sizes}}&\multicolumn{2}{l}{\textbf{Avg Length}}&  \multicolumn{3}{l}{\textbf{Benign TLDs}}&\multicolumn{3}{l}{\textbf{Malicious TLDs}}\\
             \cmidrule(lr){2-4} 
             \cmidrule(lr){5-6}
            \cmidrule(lr){7-9}
             \cmidrule(lr){10-12}
            &malicious\tnote{5}  &benign  &total  &malicious &benign &.com  &ccTLDs  &others    &.com  &ccTLDs  &others \\
            \midrule
            \textbf{GramBeddings\tnote{1}} &400,000  &400,000  &800,000& 86.24& 46.38  &52.17\%  &12.04\%  &35.79\%  &60.10\%  &11.82\%  &28.08\%   \\
            \textbf{Mendeley\tnote{2}}&35,315  &1,526,619  &1,561,934& 37.15& 35.82  &61.97\%  &0.93\%  &37.10\%  &72.86\%  &1.61\%  &25.53\%   \\
            \textbf{Kaggle 1}\tnote{3}&316,251  &316,252  &632,503 & 64.68& 58.30 &77.46\%  &0.63\%  &21.92\%  &50.59\%  &10.61\%  &38.8\%   \\
            \textbf{Kaggle 2}\tnote{4}&213,037  &428,079  &641,116& 64.13& 57.69  &74.27\%  &6.61\%  &19.12\%  &46.62\%  &7.74\%  &45.65\%   \\
            \bottomrule
        \end{tabular}
        \begin{tablenotes}[flushleft]
            \item[1,2] \label{fn:binary} These are used for binary classification, download using \href{https://web.cs.hacettepe.edu.tr/~selman/grambeddings-dataset/}{GramBeddings} and \href{https://data.mendeley.com/datasets/gdx3pkwp47/2}{Mendeley} links.
            \item[3] \label{fn:kaggle1} This is used for binary cross dataset test, download using \href{https://www.kaggle.com/datasets/samahsadiq/benign-and-malicious-urls}{this} link.
            \item[4] \label{fn:kaggle2} This is used for multiple classification, download using \href{https://www.kaggle.com/datasets/sid321axn/malicious-urls-dataset}{this} link.
            \item[5] \label{fn:malicious} Indicates malicious URLs in binary test and the total of malicious, defacement, and phishing URLs in multiple test.
        \end{tablenotes}
         \label{tab:dataset}
    \end{threeparttable}
\end{table*}

\subsection{Transformer-based Detection}
Chang et al. \cite{chang2021research} fine-tuned a BERT model, initially pretrained on English text, using URL data for detecting malicious URLs. URLTran \cite{maneriker2021urltran} comprehensively analyzed transformer models for phishing URL detection, demonstrating their feasibility and exploring various hyperparameter settings. However, these approaches, due to limited technical modifications, could not overcome the bias between the pretrained data domain and the task domain. The study in \cite{wang2023lightweight} employed a Transformer with a hybrid expert network for URL classification. Xu et al. \cite{xu2021transformer} used a lightweight Transformer-based model. Although these methods achieved good performance, they did not fully leverage the advantages of pretraining. In the work of Wang et al. \cite{wang2023large}, a domain-specific BERT architecture was pretrained from scratch for URL applications. While this approach offers many benefits, it requires extensive URL data, substantial computational resources, and extensive training time.

\blue{Compared to pre-trained models such as BERT and RoBERTa, which also use transformer architectures, TransURL exhibits superior performance in capturing character-level information and local patterns crucial for malicious URL detection. This advantage stems from its dual-channel architecture, addressing the limitations of pre-trained models that rely solely on token-based input mechanisms. Additionally, TransURL incorporates multi-scale pyramid features, enabling the analysis of URLs at various granular levels to detect patterns and anomalies indicative of malicious activity. Traditional pre-trained models generally lack such mechanisms for multi-scale analysis, which can constrain their effectiveness. Moreover, while pre-trained models employ general attention mechanisms effective for text comprehension and generation, these mechanisms are not optimized for the unique structural characteristics of URLs. TransURL's specialized attention mechanisms enhance its ability to discern intricate patterns and structural relationships within URL sequences, thereby improving detection performance.}

\section{Large Scale URL Dataset}
\label{sec:data}

The four datasets used for training, validation and testing are publicly available. These datasets share a similar schema, consisting of the browsing URL and a corresponding label indicating whether the URL has been identified as malicious or benign. Upon analyzing their statistical data, including sample size, average URL length, and top-level domain (TLD) types, as shown in \blue{Table \ref{tab:dataset}}, we discovered certain variations in the URL data across these datasets, which contribute to a more comprehensive evaluation of our model.

\textbf{GramBeddings Dataset:} As provided by GramBeddings\cite{bozkir2023grambeddings}, this dataset comprises 800,000 samples, equally divided into 400,000 malicious and 400,000 benign URLs. Malicious URLs were collected from websites such as PhishTank and OpenPhish, spanning the period from May 2019 to June 2021. Long-Term and Periodical Sampling, as well as Similarity Reduction techniques, were applied to the malicious data. The benign URLs were iteratively crawled from Alexa and the top 20 most popular websites in 20 different countries, and then randomly sampled. As a result, this dataset presents the highest diversity and sample size compared to others, while also demonstrating an equal number of instances per class.  As shown in \blue{Table \ref{tab:dataset}}, the average length of malicious URLs is significantly longer than that of benign URLs, approaching twice the length, ensuring the similarity of the class-level average length distribution. In terms of top-level domain (TLD) features, because different domains are generally more difficult to obtain in malicious URLs, this dataset has improved the domain-level diversity of malicious URLs by setting a low ratio of unique domains to total domains, achieving a ratio similar to that of benign URLs, with \textit{.com} at 60.10\% and ccTLDs \blue{(country code top-level domain)} at 11.82\%, whereas for benign URLs, it is \textit{.com} at 52.17\% and ccTLDs at 12.04\%.

\textbf{Mendeley Dataset:} From Mendeley Data \cite{singh2020malicious}, this dataset consists of 1,561,934 samples, with a significant skew towards benign URLs (1,526,619) compared to 35,315 malicious URLs. The samples were crawled using the MalCrawler tool and validated using the Google Safe Browsing API\cite{safe-browsing}. This dataset exhibits a notable class imbalance at a ratio of approximately 1:43. In terms of average length, both malicious and benign URLs demonstrate similar values, and the diversity of top-level domains (TLDs) is limited, primarily concentrated in \textit{.com}, accounting for 61.97\% and 72.86\%, respectively, with ccTLDs representing only 0.93\% and 1.61\%. Although this may pose a risk of misleading the training processes by capturing inadequate syntactical or semantic features, considering the potential encounter with such data distribution in real-world scenarios, we chose to employ this dataset for model evaluation.

\textbf{Kaggle Dataset:} 
The Kaggle 1 and Kaggle 2 datasets are both derived from the Kaggle website. Kaggle 1 is designed for binary classification experiments, while Kaggle 2 is intended for multi-classification tasks. Kaggle 1 consists of 632,503 samples, evenly distributed between malicious and benign URLs. In comparison to the Mendeley dataset, this dataset demonstrates disparity in the average length of malicious and benign URLs, with the samples presenting a more balanced composition between the two classes. Notably, within the malicious samples of this dataset, there is a noticeable higher ratio of ccTLDs in TLDs compared to the benign URLs, accounting for 10.61\%, while benign URLs only make up 0.63\%. Additionally, the proportion of \textit{.com} domains is 50.59\% for malicious URLs, whereas it is 77.46\% for benign URLs.

The Kaggle 2 dataset consists of four classes: benign (428,079), defacement (95,306), phishing (94,086), and malicious (23,645). The benign class contains positive samples, while the other three classes contain negative samples of different types. We observe that the \textit{.com} TLDs are dominant in benign URLs, accounting for 74.27\% of the total. The ccTLDs are slightly more frequent in this dataset (6.61\%) than in the other two datasets, while the other gTLDs \blue{(generic top-level domain)} represent 19.12\% of the benign URLs. For the negative samples, the \textit{.com} TLDs are less prevalent, with a frequency of 46.62\% across all three classes. The ccTLDs and other gTLDs have higher frequencies of 7.74\% and 45.65\%, respectively, in the negative samples than in the benign ones.

The distinctive composition and TLD distribution in each dataset offer a comprehensive foundation for assessing the efficacy of our method across diverse web domains. This enables robust testing under different real-world scenarios.

\section{Methodology}
\label{sec:method}

The proposed method employs the CharBERT \blue{(Character-aware Pre-trained Language Model)} network structure as the backbone network, integrating an encoder feature extraction module, a multi-scale learning module, and a  Spatial Pyramid Attention module.

The overall model structure is depicted in Fig. 3.
\begin{itemize}
    \item \textbf{Backbone Network}: The CharBERT model improves URL data interpretation and analysis with its advanced subword and character-level embedding which extends based on BERT.
    \item \textbf{Encoder Feature Extraction}: This module extracts multi-layer encoder features from CharBERT, aiming to capture URL representations ranging from low-level to high-level.
    \item \textbf{Multi-scale Learning}: 
The module conducts multi-scale local information extraction from multi-layer encoder features and captures the relational information between different encoder feature layers.
    \item \textbf{Spatial Pyramid Attention}: 
This module differentially weights different regions of the feature, highlighting local spatial correlations, allowing flexible focus on information-rich segments in URLs. This contrasts with Transformer's Multi-Head Attention, which prioritizes positional relationships.
    
\end{itemize}

\begin{figure}[t]
    \centering
    \includegraphics[scale=0.4]{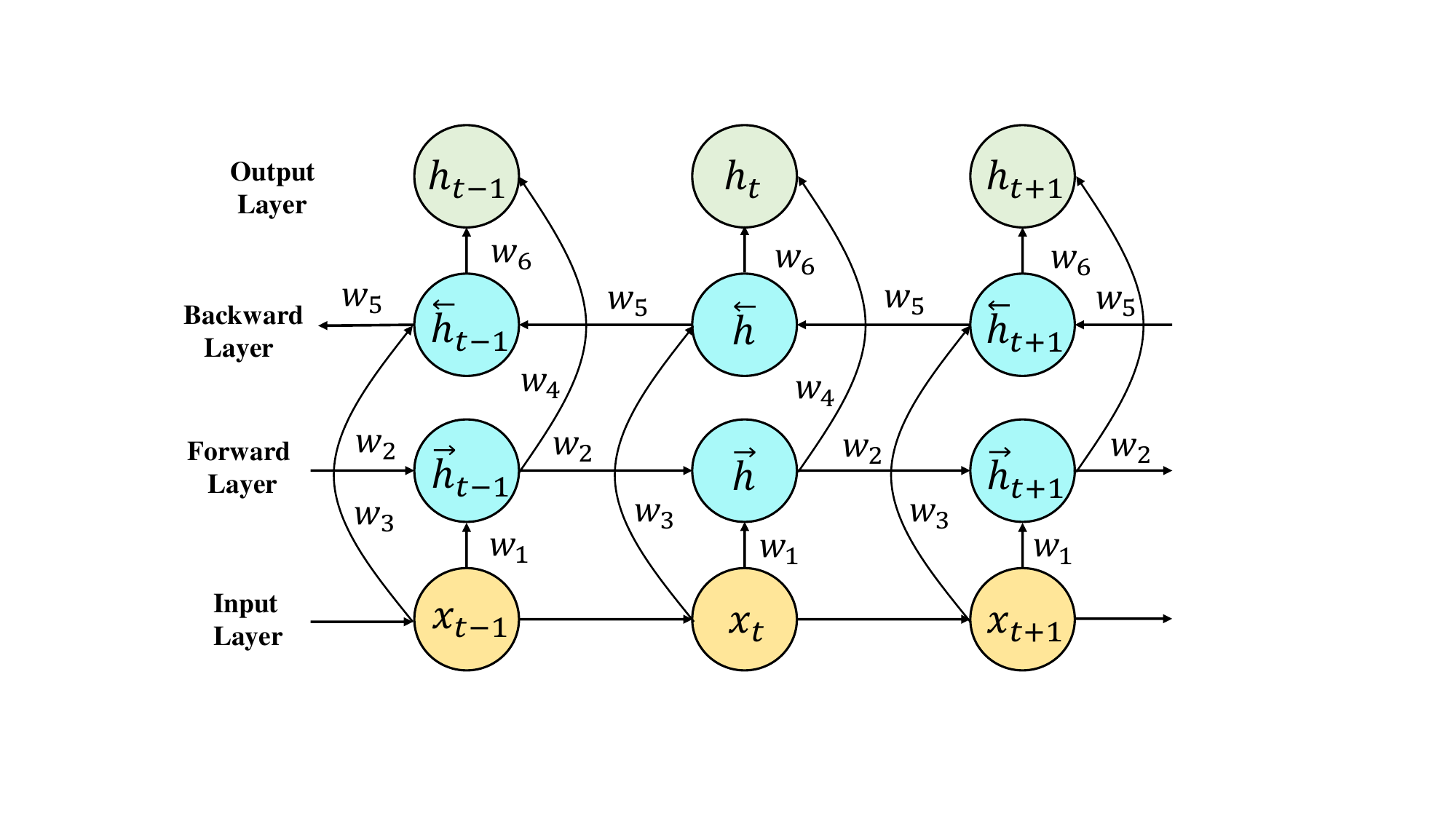}
    \caption{\blue{\textbf{A Network Structure Diagram of the BiGRU Module.}}}
    \label{fig:fig2}
\end{figure}
\begin{figure*}[t]
    \centering
    \includegraphics[width=\textwidth]{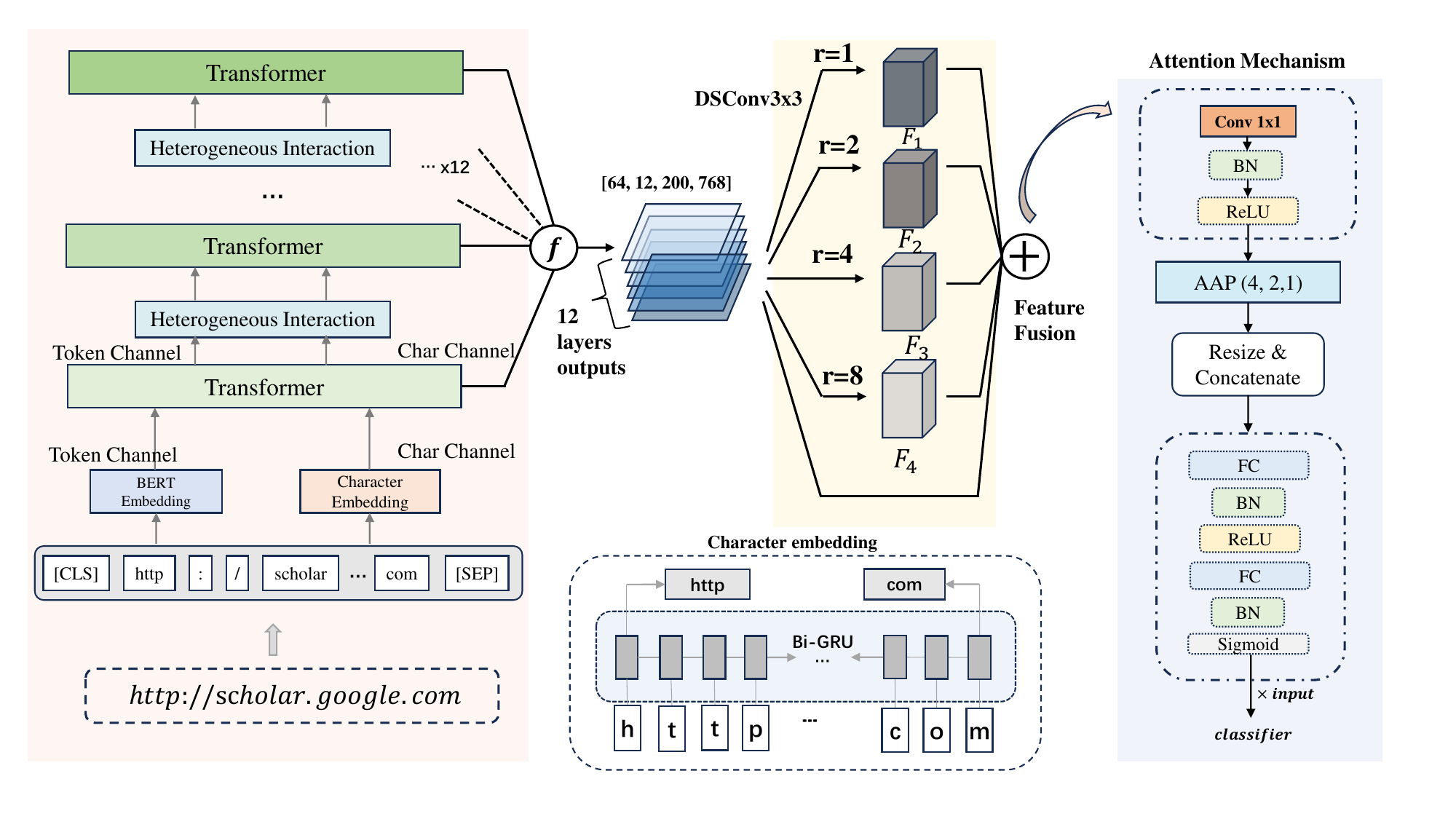}
    \caption{\textbf{TransURL: Composed of Four Core Components. CharBERT, the backbone network for learning character and subword-level features; Encoder Feature Extractor, Multi-Scale Feature Learning, and Spatial Pyramid Attention module for acquiring multi-order, multi-scale, attention-weighted features.}}
    \label{fig:fig3}
\end{figure*}

\subsection{Backbone Network}
We utilize CharBERT\cite{ma2020charbert}, an extension of BERT integrating the Transformer architecture with a dual-channel framework, as our pretrained backbone network to capture both subword and character-level features.
CharBERT capture character-level information in token sequences through two modules: (1)  the Character Embedding Module, encoding character sequences from input tokens and (2) the Heterogeneous Interaction Module, combines features from both character and subword channels, and then independently separates them into distinct representations as input for the encoder layer.

The character-aware embedding of each token is primarily generated through two components: the encoding of individual characters and subword units. These two components are integrated via a dual-channel architecture. To establish contextual character embeddings, we utilize a bidirectional Gated Recurrent Unit (BiGRU) layer. The BiGRU employs a bidirectional recurrent neural network with only the input and forget gates\cite{deng2019sequence}. The architecture diagram of the BiGRU is depicted in Fig. \ref{fig:fig2}.

Assuming $x$ denotes the input data, and $h$ represents the output of GRU unit. $r$ is the reset gate, and $z$ is the update gate. $r$ and $z$ decide how to get the new hidden state $h_t$ from the previous hidden state $h_{t-1}$ calculation. The update gate controls both the current input $x_t$ and the previous memory $h_{t-1}$, and outputs a numerical value $z_t$ between 0 and 1. The calculation formula is as follows:

\begin{equation}
z_t=\sigma(W_z[h_{t-1}, x_t]+b_x)
\end{equation}

where $z_t$ determines the extent to which $h_{t-1}$ should influence the next state, $\sigma$ is the sigmoid activation function, $W_z$ is the update gate weight, and $b_z$ is the bias. The reset gate regulates the influence of the previous memory $h_{t-1}$ on the current memory $h_t$, removing it if deemed irrelevant.

\begin{equation}
r_t=\sigma(W_r[h_{t-1}, x_t]+b_x)
\end{equation}

Then creating new memory information $h_{t}$ using the update gate:
\begin{equation}
\widetilde{h_t}=tanh(W_h[r_{t}h_{t-1}, x_t]+b_h)
\end{equation}
The output at the current moment can be obtained:
\begin{equation}
{h_t}=(1-z_t)h_{t-1}+z_{t}\widetilde{h_t}
\end{equation}

The current hidden layer state of the BiGRU is influenced by the current input $x_t$, the forward hidden state $\overset{\xrightarrow{}}{h_{t-1}}$, and the output $\overset{\xleftarrow{}}{h_t}$ of the reverse hidden layer state:

\begin{equation}
\overset{\xrightarrow{}}{h_t}=\overset{\xrightarrow{\hspace{0.5cm}}}{GRU}(\overset{\xrightarrow{\hspace{0.3cm}}}{h_{t-1}},x_t)(t = 1, 2, …, d) 
\end{equation}
\begin{equation}
\overset{\xleftarrow{}}{h_t}=\overset{\xleftarrow{\hspace{0.5cm}}}{GRU}(\overset{\xleftarrow{\hspace{0.3cm}}}{h_{t+1}},x_t)(t=d,d-1,..,1) 
\end{equation}
\begin{equation}
h_t= w_{t}\overset{\xrightarrow{}}{h_t} + v_{t}\overset{\xleftarrow{}}{h_t} + b_t = BiGRU(x_t)
\end{equation}

The GRU represents the nonlinear transformation of the input, incorporating the degradation indicator into the associated GRU hidden state. $w_t$ and $v_t$ denote the weights of the forward hidden layer state $\overset{\xrightarrow{}}{h_t}$ and reverse hidden state output $\overset{\xleftarrow{}}{h_t}$ of the bidirectional GRU at time  $t$, respectively, $b_t$ represents the bias corresponding to the hidden state at time $t$.

In the generation of character embeddings, we represent an input sequence as ${w_1,...,w_i,...,w_m}$,, where $w_i$ is a subword tokenized using Byte Pair Encoding (BPE), and $m$ is the length of the sequence at the subword level. Each token $w_i$ consists of characters ${c^{i}_{1},...,c^{i}_{n_i}}$, where $n_i$ represents the length of the subword. The total character-level input length is denoted as $N = \sum\limits_{i=1}^{m} n_i$,  where $m$ is the number of tokens. The formulation of the processing is as follows:
\begin{equation}
e^{i}_{j}=W_c \cdot c^{i}_{j};\;h^{i}_{j}=BiGRU(e^{i}_{j})
\end{equation}
Here, $W_c$ is the character embedding matrix, and $h^{i}_{j}$ denotes the representation of the $j$-th character within the $i$-th token. The BiGRU processes characters across the entire input sequence of length $N$ to generate token-level embeddings. Then connect the hidden states of the first and last characters in each token, as follows:

\begin{equation}
h_i(x)=[h^{i}_{1}(x);h^{i}_{n_i}(x)]
\end{equation}
Let $n_i$ be the length of the $i$-th token, and $h_i(x)$ be the token-level embedding from characters, enabling contextual character embeddings to capture complete word information.
\begin{figure}[t]
    \centering
    \includegraphics[scale=0.5]{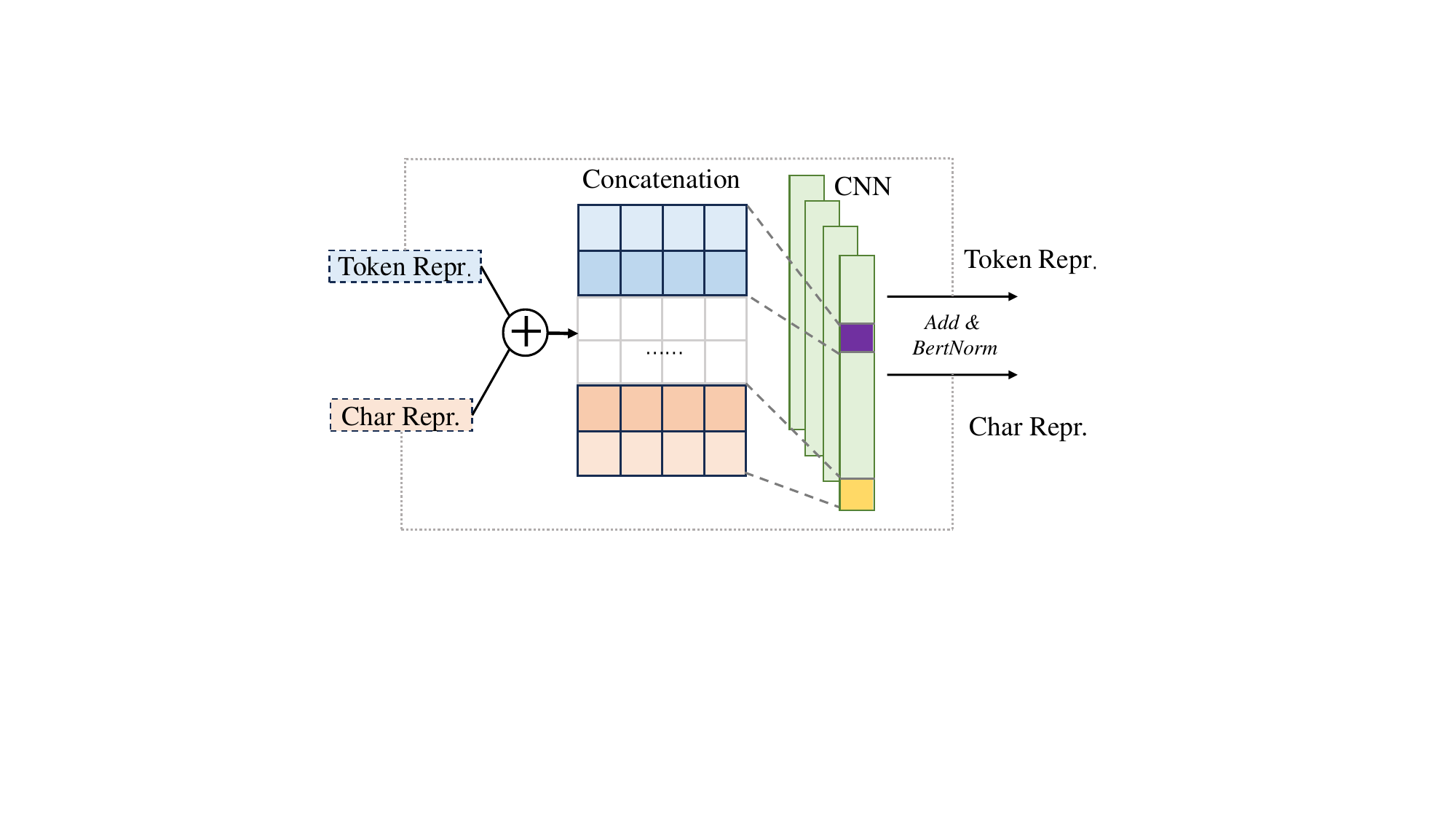}
    \caption{\textbf{The architecture of Heterogeneous Interaction Module.}}
    \label{fig:fig4}
\end{figure}

The heterogeneous interaction module fuses and separates the token and character representations after each transformer layer. The structure shown in \blue{Fig}. \ref{fig:fig4}. This module uses different fully-connected layers to transform the representations, and then concatenates and integrates them by using a CNN layer, as follows:
\vspace{10pt}
\begin{equation}
t^{'}_{i}(x)=W_1*t_i(x)+b ; \; h^{'}_{i}(x)=W_2*h_i(x)+b_2
\end{equation}
\begin{equation}
w_i(x)=[t^{'}_{i}(x);h^{'}_{i}(x)] ; \; m_{j,t}=tanh(W^{j}_{3*w_{t:t+s_j-1}}+b^{j}_3)
\end{equation}
where $t_i(x)$ is the token representation, $W$ and $b$ are the parameters, $w_{t:t+s_j-1}$ is the concatenation of the embeddings of $(w_t,...,w_{t+s_j-1})$, $s_j$ is the window size of the $j$-th filter, and $m$ is the fused representation, which has the same dimension as the number of filters.

\blue{Next is a fully connected layer with GELU activation \cite{hendrycks2016gaussian}, used to map the fused features onto two channels. A residual connection is added to preserve the original information of each channel.}

\begin{equation}
m^{t}_{i}(x)=\Delta(W_4*m_i(x)+b_4); \; m^{h}_i(x)=\Delta(W_5*m_i(x)+b_5)
\end{equation}
\begin{equation}
T_i(x) = t_i(x)+m^{t}_{i}(x); \; H_i(x)=h_i(x)+m^{h}_{i}(x)
\end{equation}
$\Delta$ is the activation function GELU, and T and H as the representations of the two channels. After the residual connection, a layer normalization operation is applyed. The fusion and separation process can enrich the mutual representations of the two channels, while preserving the specific features of the tokens and characters. The pre-training tasks can also enhance the differentiation of the dual-channel framework.

\subsection{Encoder Feature Extraction}
Pre-trained language models such as BERT use multiple layers of Transformer encoders to learn semantic knowledge from large-scale corpora, and then fine-tune them for specific downstream tasks. Most BERT-based classification models depend on the [CLS] feature of the final layer, which summarizes the semantic information of the whole input sequence. However, Jawahar \etal \cite{jawahar2019does} show that BERT can learn various information across layers, such as phrase-level details in lower layers, syntactic information in middle layers, and rich semantic features in higher layers. They apply k-means clustering to the BERT layer representations and measure the cluster quality by using Normalized Mutual Information (NMI). They find that lower BERT layers are better at encoding phrase-level information, as indicated by higher NMI scores\cite{jawahar2019does}, as shown in Table \ref{tab:tab3}. Deeper encoder layers are more effective in handling long-range dependency information.

Although each layer in the BERT family of models takes the output features of the previous layer as input for computation, multiple intricate calculations within each layer's processing may still result in potential degradation of lower-level and mid-level features, which is detrimental to the complete feature learning process. Li \etal \cite{Li2020} use feature concatenation to integrate aspect features from each layer of BERT for aspect term sentiment classification. This approach, instead of relying only on the final layer for classification features, effectively enhances classification performance by leveraging the distinct features learned at each layer of BERT.

\begin{table}[t]
    \centering
    \setlength\tabcolsep{7pt}\caption{\textbf{\blue{Performance of span representation clustering derived from various layers of CharBERT.}}}
    \scalebox{1.1}{
    \begin{tabular}{cccccccccccccc}
    \toprule
         Layer & 1 & 2 & 3 & 4 & 5 & 6 & 7 & 8 & 9 & 10 & 11 & 12 \\
         \midrule
         NMI & 0.38 & 0.37 & 0.35 & 0.3 & 0.24 & 0.2 & 0.19 & 0.16 & 0.17 & 0.18 & 0.16 & 0.19 \\
    \bottomrule
    \end{tabular}}
    \label{tab:tab3}
\end{table}

Similar to prior research, we extract outputs from each encoding layer in CharBERT. However, instead of concatenating these layer-wise features, we reorganize them into a higher-dimensional matrix. In this restructured feature matrix, layers function akin to channels in an image. The feature process is as follows: Consider a sequence of outputs $k_1, k_2, ..., k_n$ and $u_1, u_2, ..., u_m$, where each output $k_i$ and $u_j$ has a rank of $(H, W, C)$, representing the outputs of CharBERT’s word-level and character-level encoders at various layers. $H$ is the batch size, $W$ is the fixed URL sequence length (200 in our model), and $C$ is a 768-dimensional vector for each merged hidden layer output in CharBERT. For example, $k_1$ and $u_2$ be two tensors representing the sequence and character embeddings, respectively. Let $w$ be the sequence length, and $d$ be the embedding dimension:

\begin{equation}
\begin{aligned}
k_1 &= \begin{bmatrix} x_{11}^1 & x_{12}^1 & \cdots & x_{1d}^1 \\ x_{21}^1 & x_{22}^1 & \cdots & x_{2d}^1 \\ \vdots & \vdots & \ddots & \vdots \\ x_{w1}^1 & x_{w2}^1 & \cdots & x_{wd}^1 \end{bmatrix}, \quad u_1&=\begin{bmatrix} x_{11}^2 & x_{12}^2 & \cdots & x_{1d}^2 \\ x_{21}^2 & x_{22}^2 & \cdots & x_{2d}^2 \\ \vdots & \vdots & \ddots & \vdots \\ x_{w1}^2 & x_{w2}^2 & \cdots & x_{wd}^2 \end{bmatrix}
\end{aligned}
\end{equation}

Afterwards, we use one-dimensional convolution to fuse the concatenated channel features, reducing their dimensionality to the original values of each channel. Here, $C$ represents concatenation, $K_{fuse}$ denotes convolution, and $Y$ is the resulting fused tensor:

\begin{equation}
Y = K_{fuse}(C(k_1,u_1)^T) = \begin{bmatrix} y_{11} & y_{21} & \cdots & y_{m1} \\ y_{12} & y_{22} & \cdots & y_{m2} \\ \vdots & \vdots & \ddots & \vdots \\ y_{1d} & y_{2d} & \cdots & y_{md} \end{bmatrix}
\end{equation}

By stacking the merged output along the new dimension 0, we form a tensor $F$ of rank $(N, H, W, C)$, where $N$ is the number of layers (12 in our model). To align the multilevel features for subsequent analytical tasks, the tensor elements were rearranged by permuting dimensions 0 and 1. This resulted in a tensor $F' = (H, N, W, C)$, which served as the stacked feature input for the next multi-attention module.
 
\subsection{Multi-scale Learning}
In extracting URL embedding features, the standard architecture of the Transformer model lacks specialized design for capturing local features. This means it can comprehend the context of the entire input sequence but may not focus on local details. However, capturing local features in URLs is crucial, especially in applications like security analysis, fraud detection, or content categorization. Local features, including specific word patterns, character combinations, or structural anomalies, can be indicative of the nature and intent of a URL. 

To more effectively capture these local features, we augment our architecture with multi-scale feature learning, specifically designed for local feature extraction. This module is based on depthwise separable convolutions (DSConv) \cite{liu2021samnet}, offering reduced floating-point operations and enhanced computational efficiency. We employ dilated convolutions with varying dilation rates to capture multi-scale information, serving as fundamental operators for expanding the network's depth and breadth. Formally, the high-dimensional feature represented by outputs from multiple encoder layers is denoted as $M \in \mathbb{R}^{C\times H\times W}$, where $C$ is the number of channels, $H$ is the height, and $W$ is the width. The process begins by applying a single DSConv (conv3 x 3) to $M$, extracting common information denoted as $F_0$ for each branch. Specifically:

\begin{equation}
\begin{aligned}
F_0 = K_0(M)
\end{aligned}
\end{equation}

$K_0$ denotes a depthwise separable conv3 x 3 operation, and dilated DSConv3x3 with different rates are applied to $F_0$ across branches ($K_i$ for branch $i$, $N$ branches). Contextual information from multiple scales is integrated through element-wise summation using a residual connection, termed:

\begin{equation}
\begin{aligned}
F_i = K_i(F_0), i=1,2,...,N
\end{aligned}
\end{equation}
\begin{equation}
F = \sum_{i=0}^{N} F_i
\end{equation}

Since the concatenation operation substantially amplifies channel count, leading to increased computational complexity and network parameters. Thus, we opt for element-wise summation. Lastly, aggregated features are reshaped using a 1 × 1 standard convolution. Formally:

\begin{equation}
\begin{aligned}
Q = K_{fuse}(F)+M
\end{aligned}
\end{equation}

$K_{fuse}$ denotes the standard conv1 x 1 operation for fusing additional information at different scales. The original feature map $M$ is integrated as a residual connection \cite{he2016deep}, aiding gradient flow and facilitating effective training. In our experiments, dilation rates of $[1,2,4,8]$ are utilized to capture contextual information at various scales.

Within the multi-scale learning module, a straightforward element-wise summation of features from various scales may inadvertently diminish the importance of informative branches while according equal significance to all scales. To mitigate this problem, we employ a spatial pyramid attention mechanism\cite{guo2020spanet}, which effectively assesses subfields across multiple scales and adjusts branch weights, enhancing the overall performance.

\subsection{Spatial Pyramid Attention}
In the multi-scale learning module, a simple element-wise summation of features from different scales may inadvertently downplay the importance of informative branches, treating all scales equally. Additionally, while Transformers offer token-level attention, for URL feature learning, regional-level attention is crucial due to the presence of distinct information-dense areas (like domain names) and regions with noise (such as random parameters) in URLs.  To address these issue, we integrate a Spatial Pyramid Attention module following our multi-scale learning \cite{guo2020spanet}.

The spatial pyramid attention mechanism comprises three key elements: point-wise convolution, spatial pyramid structure, and a multi-layer perceptron. The point-wise convolution aligns channel dimensions and consolidates channel information. The spatial pyramid structure incorporates adaptive average pooling of three different sizes, promoting structural regularization and information integration along the attention path. Multi-layer perceptron then extracts an attention map from the output of the spatial pyramid structure.

To be sepecific, we denoted adaptive average pooling and fully connected layer as $P$ and $F_{fc}$ respectively. The concatenation operation is represented as $C$, $\sigma$ denotes the Sigmoid activation function, while $R$ is referred to as resizing a tensor to a vector. The fused feature map after the multi-scale learning moule can be denoted as $Q \in \mathbb{R}^{C\times H\times W}$, the attention mechanism learns attention weights from the input and multiplies each channel in it by learnable weights to produce an output. The output of the spatial pyramid structure $S(Q)$ can be presented as:

\begin{equation}
\begin{aligned}
S(Q) = C(R(P(Q,4)),R(P(Q,2)),R(P(R,1)))
\end{aligned}
\end{equation}

Omitting the batch normalization and activation layers for the sake of clarity, the fundamental transformation $\sigma$ can be expressed as:

\begin{equation}
\begin{aligned}
\zeta(Q) = \sigma(F_{fc}(F_{fc}(S(Q))))
\end{aligned}
\end{equation}

In our experiments, the channel number $C$ is 12 according to the former process and  we adopt 3-level pyramid average pooling. In the concluding phase of our network, we apply Mean Pooling to the weighted feature map along the fixed sequence length dimension. This outcome is then integrated with a dropout layer, followed by a fully connected layer that converts URL features into a binary class representation for prediction.

\begin{table}[t]
    \centering
    \footnotesize
    \setlength\tabcolsep{10pt}
    \caption{\textbf{Detection results of \blue{TransURL} vary with the number of layers employed.}}
    \begin{tabular}{cccccc}
    \toprule
         \ layers(count)& Accuracy &Precision  &Recall  &F1-score  &AUC \\
    \midrule
         2&  0.9772&  0.9811&  0.9730&  0.9771& 0.9959 \\
         3&  0.9837&  0.9873&  0.9799&  0.9863&  0.9963\\
         4&  0.9856&  0.9917&  0.9792&  0.9854&  0.9938\\
         5&  0.9860&  0.9861&  0.9858&  0.9859&  0.9998\\
         \textbf{12}&  \textbf{0.9915}&  \textbf{0.9949}&  \textbf{0.9880}&  \textbf{0.9914}&  \textbf{0.9965}\\
    \bottomrule
    \end{tabular}
    \label{tab:tab4}
\end{table}

\section{Experiments}
\label{sec:experiments}
This section presents the detailed experimental setup and results to assess the effectiveness of our proposed method and compare it with baselines. Our experiments are primarily divided into the following components:

\begin{figure*}
    \centering
    \includegraphics[width=\textwidth]{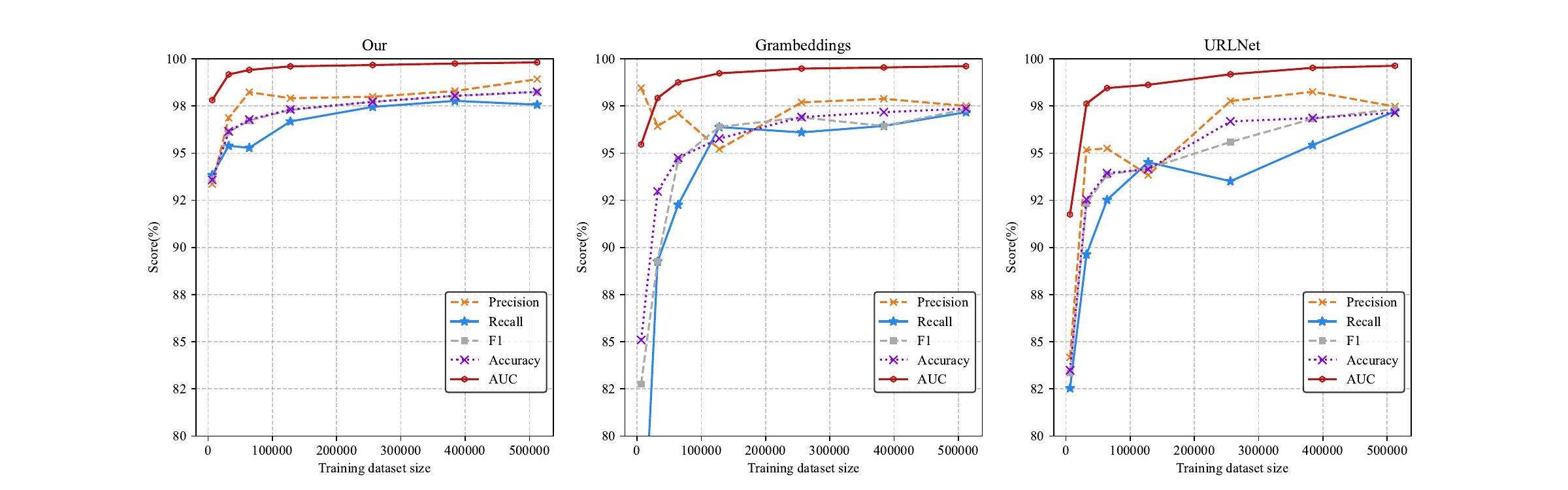}
    \caption{\textbf{Detection results of baseline methods and \blue{TransURL} on GramBeddings dataset.}}
    \label{fig:fig5}
\end{figure*}

\begin{enumerate}
    \item Assess stacked feature layer configurations for multi-layer extraction.
    \item Explore data scale dependency with varied training dataset sizes.
    \item Evaluate model generalization across datasets.
    \item Multi-class classification.
    \item Test robustness using adversarial samples.
    \item Evaluate practicality with recent active malicious URLs.
\end{enumerate}

\textbf{Setup.} The pre-trained CharBERT was trained on the English Wikipedia corpus, consisting of a total of 12GB and approximately 2,500 million words. Hyperparameter tuning during fine-tuning led to a batch size of 64, AdamW optimizer (initial learning rate: 2e-5, weight decay: 1e-4), 0.1 dropout rate, and 5 training epochs. We utilized PyTorch 2.0, NVIDIA CUDA 11.8, and Python 3.8, conducting training on NVIDIA A100 GPUs. The final model for each experiment was selected based on the best validation loss.

\textbf{Baselines.} In our experimental comparison, we chose the state-of-the-art models, URLNet and GramBeddings, as benchmarks for evaluating our proposed method. To ensure fairness and reproducibility, we obtained their code from GitHub repositories without making any modifications to the structure or hyperparameters, and applied them to our dataset. Specifically, for URLNet, we selected Embedding Mode 5, the most complex mode, as it exhibited superior performance in their original study.

\subsection{Evaluate the Effectiveness of Multi-level Features}
We develop a multi-layer feature extraction module to distill semantic features from different layers of the Transformer encoders in our backbone network. Our approach involves stacking embedding outputs from different layers to form a multi-layered feature matrix, which we optimize to enhance URL semantic feature representation. Through this, we investigate the complexity of feature representation across the network and aim to demonstrate the effectiveness of multi-layered features in improving overall model performance.

We create a training corpus from the GramBeddings dataset, randomly selecting 128k URLs from the 80k available in the training set while maintaining the proportion of malicious and benign URLs. We also randomly sample 32k URLs for testing and validation respectively. The performance of various stacked configurations and the incremental gains are shown in Table \ref{tab:tab4}. We observe a consistent improvement in evaluation metrics as we incrementally incorporate additional layers, namely the last 2, 3, 4, and 5 layers. Ultimately, the stacking of 12 embedding output layers achieves the best performance in URL detection tasks, demonstrating the effectiveness of integrating both lower and deeper layers in the model architecture.

\subsection{Comparison with Baselines}
In this section, we compare the performance of TransURL with baselines using binary and multi-class malicious URL detection.
\begin{table*}[t]
    \centering
    \caption{\textbf{Detection results of baseline methods and \blue{TransURL} on Mendeley dataset.}}
    \begin{tabular*}{0.85\textwidth}{@{\extracolsep{\fill}}l@{\hspace{0.85cm}}l@{\hspace{0.8cm}}c@{\hspace{0.6cm}}c@{\hspace{0.6cm}}c@{\hspace{0.6cm}}c@{\hspace{0.6cm}}c}
    \toprule
         Training Size & Method & Accuracy & Precision & Recall & F1-score & AUC \\
    \midrule
         \multirow{3}{*}{629,184 (60\%)} & URLNet &0.9858  &0.9475  &0.3889  &0.5515  &0.9046 \\
         & GramBeddings &0.9801  &0.6137  &0.3026  &0.4053  &0.8205 \\
         & \textbf{\blue{TransURL}} &0.9886  &0.9104  &\textbf{0.5419}  &\textbf{0.7027}  &\textbf{0.9438} \\
    \midrule
         \multirow{3}{*}{419,456 (40\%)}&URLNet &0.9842  &0.9810  &0.3019 &0.4617 &0.8992\\
         & GramBeddings  &0.9794  &0.9984  &0.0817 &0.1510 &0.8750 \\
         & \textbf{\blue{TransURL}} &0.9882  &0.8998  &\textbf{0.5307}  &\textbf{0.6677}  &\textbf{0.9370} \\
    \midrule   
         \multirow{3}{*}{209,064 (20\%)} &URLNet &0.9785  &0.9653  &0.0450  &0.0860 &0.7762 \\
         &GramBeddings  &0.9804  &0.9677  &0.1306  &0.2301  &0.7869 \\
         & \textbf{\blue{TransURL}} &0.9879  &0.9131  &\textbf{0.5055}  &\textbf{0.6507}  &\textbf{0.9322} \\
    \midrule   
         \multirow{3}{*}{104,832 (10\%)} &URLNet &0.9789  &0.8382  &0.0735  &0.1351 &0.7584 \\
         &GramBeddings  &0.9757  &0.3879  &0.1423  &0.2082  &0.8153\\
         & \textbf{\blue{TransURL}} &0.9865 &\textbf{0.8540} &\textbf{0.4774}  &\textbf{0.6125}  &\textbf{0.9161} \\
    \midrule   
         \multirow{3}{*}{104,832 (5\%)} &URLNet &0.9776  &0.0000  &0.0000  &0.0000 &0.6746 \\
         &GramBeddings  &0.9782  &0.5647  &0.1139  &0.1896  &0.7752\\
         &\textbf{\blue{TransURL}} &0.9861 &\textbf{0.8706} &\textbf{0.4397}  &\textbf{0.5843}  &\textbf{0.8994} \\
    \midrule   
         \multirow{3}{*}{10,432 (1\%)} &URLNet &0.9776  &0.0000  &0.0000  &0.0000 &0.4419 \\
         &GramBeddings  &0.9722  &0.1808  &0.0682  &0.0991  &0.6185\\
         & \textbf{\blue{TransURL}} &0.9834&\textbf{0.7632} &\textbf{0.3717}  &\textbf{0.5000}  &\textbf{0.8550} \\
    \bottomrule
    \end{tabular*}
    \label{tab:tab5}
\end{table*}

\begin{table*}[t]
    \centering
    \caption{\textbf{Cross-dataset performance generalization.}}
    \begin{tabular*}{\linewidth}{@{\extracolsep{\fill}}c@{\hspace{0.5cm}}l@{\hspace{0.5cm}}c@{\hspace{0.5cm}}c@{\hspace{0.5cm}}c@{\hspace{0.5cm}}c@{\hspace{0.5cm}}c}
       \toprule
         Cross-dataset & Method & Accuracy & Precision & Recall & F1-score & AUC \\
    \midrule
        \multirow{3}{*}{Gram/Kaggle} & URLNet  & 0.8823  & 0.8947  & 0.8666  & 0.8804  & 0.9492 \\
         & GramBeddings  & 0.5214  & 0.5120  & 0.8595  & 0.6552  & 0.4647 \\
         & \textbf{\blue{TransURL}}  & \textbf{0.9138}  & \textbf{0.9576}  & \textbf{0.8641}  & \textbf{0.9085}  & \textbf{0.9705} \\
    \bottomrule
    \end{tabular*}
    \label{tab:tab6}
\end{table*}
\subsubsection{Binary classification}

In the binary classification detection task, we used two datasets with significant differences. The first is the GramBeddings dataset, characterized by a balanced distribution of positive and negative samples and high diversity. The second is the Mendeley dataset, which exhibits extreme class imbalance and lower diversity. These datasets are used for evaluations in different detection scenarios. We explored the model's dependency on training data size by varying it, starting from as low as 1\%. Specifically, the experimented training sizes include 1\% , 5\%, 10\% , 20\%, 40\% , 60\% , and 80\% . And the trained models are tested across all the test datasets.

\textbf{Results on GramBeddings dataset:} As shown in Fig. \ref{fig:fig5}, our proposed method achieves superior performance over the baseline method on the GramBeddings dataset, regardless of the size of the training set. Remarkably, our model demonstrates high proficiency even with scarce training data.

It is worth noting that, our method achieves remarkable performance with only 6,400 URLs (1\%) for training, attaining an accuracy of 0.9358, which surpasses the baseline methods that range from 0.8349 to 0.8509. Furthermore, our model exhibits a high sensitivity in detection, with a recall of 0.9384, compared to the baseline recall of 0.7131 and 0.8254. The maximum gap in F1 score reached 0.1084.

Despite the gradual improvement in the baseline model's performance with larger training samples, narrowing the gap with our method, our approach consistently achieves an accuracy of 0.9825 and an F1 score of 0.9824 using 80\% of the training dataset. Our model outperformed the best baseline model with an accuracy and F1 score 0.0089 and 0.0091 higher. Although the difference seems small, it becomes significant when dealing with large-scale datasets. In conclusion, our model exhibits superior performance in accurately detecting malicious URLs on a balanced dataset, even with a small training set size.

\textbf{Results on Mendeley dataset:} To evaluate our method in real-world internet scenarios, where phishing sites are significantly outnumbered by legitimate web pages, we use the Mendeley dataset for further testing. This dataset contains 1,561,934 URLs, with a high imbalance ratio of about 43 to 1 between benign and malicious samples. This extreme imbalance poses a notable challenge to model performance, as it may cause a bias towards the abundant benign URL samples and increase the false positive rate when detecting malicious samples.

As shown in Table \ref{tab:tab5}, \blue{TransURL} exhibits significant advantages in class-imbalanced scenarios compared to other approaches. With just 1\% of training data, \blue{TransURL} achieves an accuracy of 0.9837, with a Precision of 0.7632, surpassing the best baseline Precision of 0.1808. The F1 score of \blue{TransURL} is four times higher than the best baseline performance. As training data increases, our F1 score reaches 0.7027, substantially exceeding the baseline peak of 0.5515, with a Recall 15.3 higher than the best baseline result. These experiments demonstrate the substantial improvements \blue{TransURL} brings in identifying malicious URL samples, significantly reducing false positives. Moreover, the high AUC \blue{(Area Under the Curve)} reflects our model's confidence in the identified samples, indicating its accurate capture of key differences between malicious and benign samples.

We observe that while URLNet achieves high accuracy on larger datasets, it suffers from a high false negative rate, particularly on smaller datasets. Notably, URLNet fails to identify any malicious URLs when the training data size is reduced to below 5\%. In contrast, GramBeddings shows significant sensitivity to data, with its performance in detecting malicious URLs varying greatly across different training sample sizes. For instance, at 60\% training data, its Precision drops to 0.6137, and at 40\%, its Recall falls to just 0.0817. Compared to these, TransURL demonstrates consistent and reliable performance across various data scales, indicating its robustness to small-scale and class-imbalanced data scenarios.

\blue{These notable performance improvement with limited training data can be attributed to several key factors. Firstly, TransURL utilizes multi-layer transformer encoding, enabling efficient capture of long-range dependencies and intricate patterns within URLs. This capability is particularly advantageous in scenarios with scarce training data, as the model demonstrates superior generalization from fewer examples. Secondly, by integrating multi-scale pyramid features, TransURL is capable of analyzing URLs at various granularities. This multi-scale approach ensures the detection of critical features at different levels, thereby enhancing the model's ability to distinguish between benign and malicious URLs even with limited training data. Moreover, we employ transfer learning techniques, wherein TransURL is pre-trained on a larger textual dataset before being fine-tuned for the specific task of URL classification. This pre-training phase equips the model with a robust foundational understanding, significantly enhancing its performance during fine-tuning with limited data.}

\subsubsection{Multi-classification:}
\begin{figure}[t]
    \centering
    \includegraphics[scale=0.5]{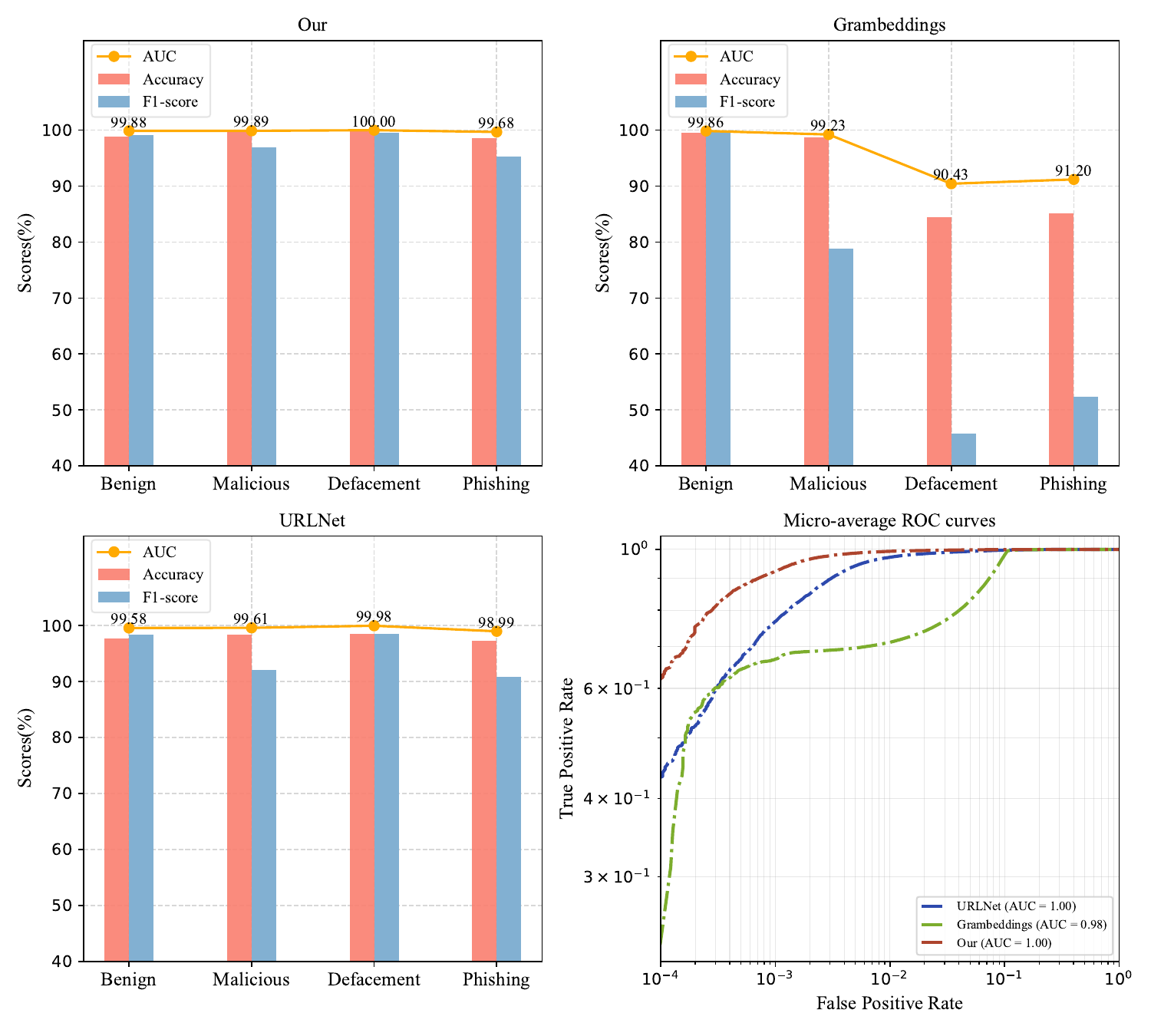}
    \caption{\textbf{Detection results of baseline methods and \blue{TransURL} on multiple classification dataset.}}
    \label{fig:fig6}
\end{figure}

To evaluate \blue{TransURL} in the context of complex cyber threats, we conduct a multi-class classification experiment, using a Kaggle 2 dataset \cite{kaggle} with four URL categories: benign (428,079), defacement (95,306), phishing (94,086), and malicious (23,645). \blue{Fig}. \ref{fig:fig6} shows the results of our model and the baseline methods.

Fig.\ref{fig:fig6} illustrates the performance of our method and the baseline methods across all four categories. \blue{TransURL} surpasses the baseline methods in each metric. The average ROC \blue{(Receiver Operating Characteristic curve)} curve highlights our method's efficacy with a TPR \blue{(True Positive Rate)} of almost 90\% at a low FPR\blue{(False Positive Rate)} of 0.001, surpassing other methods that achieve around 75\%. GramBeddings struggled in recognizing negative samples from various categories, resulting in a 50\% F1 score for defacement and phishing URLs and an overall accuracy of 83.91\%. URLNet achieved an overall accuracy of 97.07\%, falling short of our model's 98.57\%. These results demonstrate the robustness and effectiveness of \blue{TransURL} in complex multi-class classification tasks, indicating its potential as a promising solution for malicious URL detection in cybersecurity.

\subsection{Cross-dataset Testing}

To evaluate the generalization of models and to identify any potential weaknesses or biases, we set up a cross-dataset testing experiment. We do this by training the model on the GramBeddings dataset and subsequently testing it on the Kaggle binary classification dataset. It is noteworthy that the GramBeddings and Kaggle datasets significantly differ in their data collection times and sources.

The results, as shown in Table \ref{tab:tab6}, indicate a marked decline in performance of baseline methods on data not included in their training set, with URLNet's accuracy dropping to 0.8823 and GramBeddings' accuracy reducing to 0.5214 and 0.4647, respectively. In contrast, \blue{TransURL} maintained high accuracy on external datasets, achieving an AUC of 0.9705. This underscores that the knowledge gained from one dataset by our method can be effectively generalized to others, even if the data was collected much later than \blue{TransURL}'s pretraining period. This demonstrates the adaptability and long-term applicability of our approach to different data environments.
\subsection{Evaluation against Adversarial Attacks}
\begin{figure}[t]
    \centering
    \includegraphics[scale=0.7]{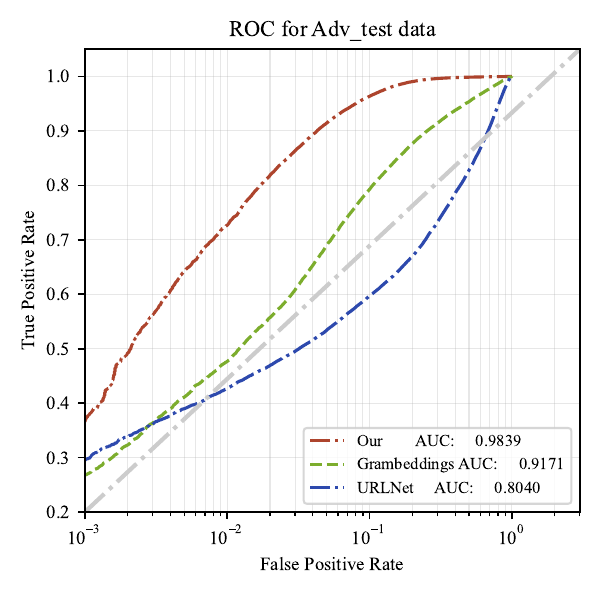}
    \caption{\textbf{Area under ROC curve under adversarial attack.}}
    \label{fig:fig7}
\end{figure}
\label{sec:eval}
 Cybercriminals employ adversarial attacks to bypass systems by exposing them to inaccurate, unrepresentative, or malicious data. We employed a Compound Attack technique as our threat model, which involves inserting an evasion character to a benign URL sample to create a real-world compatible malicious URL. This technique was first proposed by Maneriker \textit{et al.}\cite{maneriker2021urltran} and later applied and extended by GramBeddings\cite{bozkir2023grambeddings}.

The generation of adversarial samples entails the utilization of XLM-RoBERTa \cite{conneau2019unsupervised} for domain tagging in provided URLs. This process ensures a minimal tag count, involves the random insertion of hyphens in split parts, and includes the substitution of benign domains with malicious ones, resulting in the creation of an adversarial list.

To construct the AdvTest set, we merged 80K legitimate URLs and randomly sampled 40K malicious samples from the original validation data. Furthermore, we introduced 40K adversarial samples generated from benign URLs. This novel dataset presents a substantial challenge to the robustness of the model. Then we evaluate our and baseline models using this novel dataset.

As illustrated in Fig. \ref{fig:fig7} and Table \ref{tab:tab7}, baseline methods experience a significant decline in performance under adversarial sample attacks. The accuracy of URLNet decreases to 76.10\%, and that of GramBeddings to 70.18\%, while our model maintains an accuracy above 90\%, with an AUC of 98.39\%, exceeding URLNet by about 20\%. At a fixed FPR of 0.01, \blue{TransURL} achieves a TPR of nearly 75\%, more than double the TPR of baseline methods, both under 30\%. These results indicate the robustness of our approach to adversarial sample attacks, suggesting increased effectiveness in preventing malicious attack evasion in real-world scenarios.
\begin{table}[t]
    \centering
    \setlength\tabcolsep{10pt}  
    \caption{\textbf{Performance under adversarial attack.}}
    \scalebox{1.0}{
    \begin{tabular}{cccccc}
        \toprule
        Method & ACC & P & R & F1 & AUC \\
        \midrule
        URLNet & 0.7610 & 0.8745 & 0.5635 & 0.6854 & 0.8040 \\
        Gram & 0.7018 & 0.6137 & 0.8564 & 0.8564 & 0.8564 \\
        \blue{TransURL} & \textbf{0.9023} & \textbf{0.9738} & 0.8104 & \textbf{0.8846} & \textbf{0.9839} \\
        \bottomrule
    \end{tabular}}
    \label{tab:tab7}
\end{table}

\begin{table*}[t]
    \centering
    \begin{threeparttable} 
        \caption{\textbf{Cross-dataset performance generalization.}}
        
        \begin{tabular}{lccc}
            \toprule
            \textbf{Malicious url} & \textbf{URLNet} & \textbf{GramBeddings} & \textbf{Our} \\
            \midrule
            https://bafybeibfyqcvrjmwlpipqkdyt2xr46cea7ldciglcbybfwtk7cieugcj3e.\\ipfs.infura-ipfs.io & \checkmark &\ding{55} &\checkmark \\
            http://798406.selcdn.ru/webmailprimeonline/index.html & \checkmark &\ding{55} &\checkmark  \\
            https://79efc264-a0d7-4661-900b-a8bc1443be89.id.repl.co/biptoken.html & \checkmark &\ding{55} &\checkmark  \\
            http://ighji.duckdns.org &\ding{55} &\ding{55} &\checkmark \\
            https://www.minorpoint.lqoipum.top/ & \checkmark &\ding{55} &\checkmark \\
            https://colstrues.com/s/jsrj &\ding{55} &\ding{55} &\checkmark\\
            https://innovativelogixhub.firebaseapp.com/ & \checkmark &\ding{55} &\checkmark\\
            https://sites.google.com/view/dejoelinoctskxo2bb &\ding{55} &\checkmark &\checkmark \\
            https://gtly.to/-H0PPiKyq &\ding{55} &\checkmark &\checkmark \\
            \bottomrule
        \end{tabular}
        \begin{tablenotes}
            \item \textbf{Note:} We use the symbols \checkmark and \ding{55} to denote the correct and incorrect classification results, respectively.
        \end{tablenotes}
        \label{tab:tab8}
    \end{threeparttable}
\end{table*}

\subsection{Case Study}
\label{sec:case}
We conduct a series of case studies applying our detection model to active malicious web pages to evaluate its practical utility. In November 2023, we crawled 30 active phishing URLs reported and verified on PhishTank and tested them with our model trained on 30\% of the GramBeddings dataset. For comparison, tests were also conducted with the best-performing URLNet and GramBeddings models, trained on the same dataset. Results indicate that \blue{TransURL} detected all malicious URLs with 100\% accuracy, while GramBeddings misclassified 7 URLs, resulting in 76\% accuracy, and URLNet misclassified 4 URLs, with an accuracy of 86\%. Table \ref{tab:tab8} lists URLs that were incorrectly classified, in order to provide a detailed perspective on the performance of each model.

URLNet failed in detecting four malicious URLs, which had relatively short strings. Given that benign URLs on the internet are typically simple, URLNet might have mistakenly classified these short malicious URLs as benign due to their length similarity. This suggests that URLNet relies excessively on URL string length for classification, demonstrating limited capability in recognizing semantics and specific patterns in real-life scenarios. Conversely, GramBeddings, which combines convolutional neural networks, long short-term memory networks, and attention layers, is a more complex system. It misclassified instances of both longer and shorter URLs, indicating a minor influence of URL length. However, the significant diversity among the seven misclassified malicious URLs implies that GramBeddings' performance could be affected by various factors, and its learning system might not have developed sufficiently generalized discriminative patterns.

In contrast, only our method demonstrated consistent or even improved performance in real-world applications, showing its adaptability to various forms of malicious URLs. Analyzing the differences at a technical level, we attribute our approach's distinct advantage over others to its comprehensive feature consideration,includeing character-aware token embeddings, multi-level and multi-scale feature learning, and regional-level attention.

\section{Discussion}
The proposed method, validated through a comprehensive set of experiments, has demonstrated robust, accurate, and reliable performance. Here, we briefly discuss the beneficial advantages and insights brought about by our proposed approach.

\begin{enumerate}
    \item \textbf{End-to-End Architecture}: TransURL is an end-to-end network utilizing pretrained CharBERT, which requires no manual feature initialization. It directly processes raw URLs and generates character-aware subword token embeddings. In contrast, previous studies typically necessitated manual initialization of character and word-level representations and relied on dual-path neural networks. Our approach streamlines the processing workflow while maintaining efficient feature extraction capabilities. 

   \item \textbf{Evaluation Metrics}: Our experiments show that TransURL significantly improves accuracy, robustness, and generalizability, consistently delivering stable and effective detection across various testing scenarios. Meanwhile, leading baseline methods, such as URLNet and Gramembedings, displayed clear weaknesses: URLNet struggled in class-imbalanced scenarios with small datasets, while Gramembedings faced significant performance fluctuation. Furthermore, these methods varied in their effectiveness in generalization and adversarial robustness tests. This highlights the need for a comprehensive performance evaluation system that goes beyond specific experimental setups.

 \item \textbf{Case Studies}: Previous research often overlooked the importance of case studies, but our work emphasizes the necessity of applying models directly to active malicious links to accurately reveal their real-world performance. Our case studies have shown that even methods excelling in experimental settings can face significant challenges in practical applications. Additionally, case studies offer an opportunity to thoroughly analyze a model's feature learning capabilities and shortcomings in information capture patterns, allowing for a more comprehensive assessment of the model's technical design.

 \blue{\item \textbf{Computational Efficiency}: A key consideration for the practical implementation of our proposed TransURL model is its computational efficiency, especially when deployed on resource-constrained devices such as endpoints. Given the intensive computational demands of Transformer-based models, the processing power required for real-time malicious URL detection can be substantial. To address this, we can employ optimization techniques such as model pruning and quantization, which significantly reduce memory usage without compromising detection accuracy. Additionally, we propose a hybrid approach where initial URL filtering is performed using a lightweight heuristic-based method. URLs flagged as potentially malicious are then subjected to more intensive scrutiny by the TransURL model. This layered approach ensures that the majority of URLs can be quickly processed with minimal computational overhead, while the TransURL model is reserved for cases where its advanced capabilities are most needed. These considerations make our approach viable for real-world applications, balancing the need for high detection accuracy with the constraints of endpoint devices. We will focus more on the solution to this problem in our future work.}

\end{enumerate}

\section{Conclusion}
\label{sec:conclusion}

We have proposed a novel transformer-based and pyramid feature learning system called TransURL for malicious URL detection. Our method effectively leverages knowledge transfer from pretrained models to URL contexts, dynamically integrates character and subword-level features, and incorporates three closely integrated feature learning modules for URL feature extraction. The key contributions of our approach are: 1) enabling end-to-end learning from raw URL strings without manual preprocessing; 2) adopting an interactive subword and character-level feature learning network architecture for improved character-aware subword representations; 3) conducting effective multi-level and multi-scale URL feature learning based on our proposed lightweight feature learning modules, addressing inherent limitations of the Transformer in local feature extraction and spatial awareness. We conduct extensive experiments on various URL datasets, demonstrating that our method consistently outperforms existing state-of-the-art baseline methods and produces stable decisions across scenarios. Furthermore, our method exhibits superior generalization and robustness in cross-dataset detection and adversarial sample attacks, enhancing its reliability in practical applications. We also provide a case study with comparative analysis to demonstrate the practical value of our method.

\section*{CRediT authorship contribution statement}
Ruitong Liu: Conceptualization, Data curation, Formal analysis, Investigation.  Yanbin Wang:  Methodology, Writing– original draft, Writing– review \& editing. Zhenhao Guo: Software, Validation, Conceptualization. Haitao Xu: Funding acquisition, Supervision. Wenrui Ma: Supervision. Fan Zhang: Project administration, Supervision.

\section*{Acknowledgements} 
The authors wish to express their sincere gratitude for the support received from the National Natural Science Foundation of China (NSFC) with the grant number 62272410. 
\bibliographystyle{splncs04}
\bibliography{main}


\end{document}